# Machine Learning in Transaction Monitoring: The Prospect of xAI

Julie Gerlings
Copenhagen Business School
Jge.digi@cbs.dk

Ioanna Constantiou
Copenhagen Business School
ic.digi@cbs.dk

**Abstract**

*Banks hold a societal responsibility and regulatory requirements to mitigate the risk of financial crimes. Risk mitigation primarily happens through monitoring customer activity through Transaction Monitoring (TM). Recently, Machine Learning (ML) has been proposed to identify suspicious customer behavior, which raises complex socio-technical implications around trust and explainability of ML models and their outputs. However, little research is available due to its sensitivity. We aim to fill this gap by presenting empirical research exploring how ML supported automation and augmentation affects the TM process and stakeholders' requirements for building eXplainable Artificial Intelligence (xAI). Our study finds that xAI requirements depend on the liable party in the TM process which changes depending on augmentation or automation of TM. Context-relatable explanations can provide much-needed support for auditing and may diminish bias in the investigator's judgement. These results suggest a use case-specific approach for xAI to adequately foster the adoption of ML in TM.*

## 1. Introduction

Reporting money laundering and terrorist financing activities is a relatively new practice that began in the 1970s. Statistical methods for detecting money laundering were not put in place until the late 1990s (Alsuwailem & Saudagar, 2020). Today, we still mainly rely on the same anti-money laundering (AML) process whereof rule-based scenarios are essential. The scenarios are general if-then rules based on results falling above or below certain thresholds, which makes them inflexible in a fluctuating world. This leads to an estimated 95-98% of false positive alerts manually handled and only estimated 5% of all money laundering and terror financing cases proceeding to authorities (Han et al., 2020). In our digitally and globally connected world, the challenges are keeping track of the increasing transaction flow and identifying new forms of illicit or illegal activities. Money laundering activities varies in shape and size, generating a scaling challenge of volume, velocity, and variety. Moreover, financial institutions risk harsh sanctions, monetary fines, and reputational damage, if they do not meet the requirements and expectations of regulators. Therefore, banks invest heavily in costly AML activities that are fast becoming insufficient for the technological advancements and growing transaction volumes, while human resources to investigate transactions are decreasing (Chen et al., 2018). Many approaches to adopting ML in AML practices have been proposed, but few reach empirical testing and implementation due to the inherent complexity of AML. With respect to transaction monitoring (TM), ML has the possibility to soften the hard-set if-then rules and take many more variables into account when evaluating a single transaction and provide a more nuanced view of a transaction. Moreover, criminal activities stem from behavioral patterns, which ML is often used to uncover. Because of the sensitivity of TM, a tendency towards using credit card fraud cases have been undertaken and the literature on ML in transaction monitoring (TM) is very sparse, fragmented, and poor in quality (Al-Suwaidi & Nobanee, 2020). Only a handful of recent papers discuss the relevant debate on the ability to provide explanations for models operating in high-risk areas such as AML/TM (Han et al., 2020; Jullum et al., 2020). The need for knowing more about how to ensure adoption, correct use of ML in TM and provide an output for understanding and validation should be possible to remain compliant. We conducted an action research project in collaboration with a European bank, over the course of two years exploring stakeholder needs and how xAI can positively affect the adoption of their transaction monitoring model (TMM). In the effort to determining information provided by xAI outputs as viable information, common struggles revolve around model accuracy vs. interpretability (Gilpin et al., 2018; Z. C. Lipton, 2018), completeness vs. interpretability (Alvarez-Melis & Jaakkola, 2018; Gilpin et al., 2018; Z. C. Lipton, 2018), trust (Shin, 2021), human vs. machine (Scantamburlo et al., 2018), objective/goal alignment from problem formulation to model output, and applicability/context creating bias in the decision-making process (Páez, 2019). The most prominent dilemmas related to the case will be addressed through the following research question: **"How can xAI assist the transition from traditional rule-based transaction monitoring to machine learning-based monitoring?"** To answer this question, the study identifies key stakeholders in the TM process involved with two models tested for automation TMM (A) and augmentation TMM (B). Thereafter, we analyze the stakeholders' information need derived from the changes in the process when testing the two different approaches. Lastly, we assess the two model objectives compared to stakeholders' information need to evaluate the usefulness of xAI frameworks.

## 2. Theoretical Background

Transaction monitoring is regarded as a highly sensitive area for financial institutions. It is extensively regulated and in high risk of sanctions and reputational damage if things go wrong. Moreover, AML represents an area of competitive advantage reflected in customer satisfaction (Canhoto, 2021; Han et al., 2020; Jullum et al., 2020). Due to the sensitivity of the topic and risk of criminals gaming the system, limited research has been done on the topic, results are fragmented and wide-ranging in quality (Canhoto, 2021; Han et al., 2020). Due to these circumstances, it is difficult for banks to share information with researchers and other groups (Canhoto, 2021; Jullum et al., 2020) in order to improve practices and generate a united front against money laundering (DFSA, 2021).

### 2.1. Money Laundering

Money laundering is defined as the process of concealing the illegal origin of any income or transaction in any form (primarily monetary funds or crypto valuta), leaving almost no visible trails besides the transaction statements in accounts at financial institutions. Money laundering is different from fraud, which is defined as the action of intentionally deceiving someone to gain a personal or financial advantage (Sabau, 2012). Detecting money laundering or terror financing transactions has become a near-impossible task for banks and regulators due to three main factors: 1) exponential growth in transactions; 2) a manual labor-intensive AML processes; and 3) a reliance on rigid, inflexible rule-based scenarios to produce an excessive number of transaction alerts for manual investigation (Jullum et al., 2020; Leo et al., 2019). Only a small percentage (~5%) of these investigations lead to SARs (Suspicious Activity Reports) handed over to governmental institutions. Here, only 10% of SARs result in investigation by law enforcement (Han et al., 2020; Weber et al., 2018). Failing to comply with AML regulations, reporting suspicious activities, and monitoring transactions have resulted in billions of dollars' worth in fines (Viswanatha & Wolf, 2012). This underlines banks' difficulties balancing legislators' requirements with the growing number of transactions in a volatile environment. In the EU, compliance includes implementation of a transaction monitoring system that screens all transactions and report suspicious activities (FATF, 2012). Moreover, banks are expected to account for technological advancements in their system, which is one of the main drivers towards ML. New elements of regulation from the Court of Justice of the European Union's (CJEU's) case law and Article 23(2) of GDPR come into play when using ML for high-stakes decisions such as TM, along with the increasing societal pressure of model monitoring and transparency to avoid the risk of models going haywire (Bertrand et al., 2021; Canhoto, 2021; European Commision, 2021; Han et al., 2020). This creates a problem that is not only a matter of automation and efficiency but also one of improving detection and effectiveness in existing systems. To address these areas, an ever-growing number of banks are considering ML solutions for assistance.

### 2.2. Transaction Monitoring and Machine Learning

AML can be seen as a two-sided approach following the FATF (Financial Action Task Force) objectives, with the first side centering around KYC (knowing your customer), customer patterns, their networks, and using this information to identify suspicious behavior (FATF, 2012). The second part of AML is centered around TM and reporting suspicious transactions, wherein ML efforts have been addressed with both supervised (known target variable) and unsupervised (pattern revelation) ML methods. This research focuses on the screening of transactions in TM based on scenario-generated alerts. The existing body of literature in this area lacks depth, concrete empirical analysis, and data foundations due to the sensitivity of the topic (Al-Suwaidi & Nobanee, 2020; Canhoto, 2021; Han et al., 2020; Jullum et al., 2020). Moreover, the spectrum of money-laundering activities is wide-ranging, creating a very broad list of ML objectives and approaches (Alsuwailem & Saudagar, 2020; Chen et al., 2018; Han et al., 2020). (FATF, 2012)There has been considerable debates on applying supervised or unsupervised methods for TM, due to the outcome of SARs rarely being known to banks. Governmental investigations can take months to reach a definitive conclusion. Therefore, supervised learners are often discarded (Jullum et al., 2020). However, the framing of the target variable is up for discussion, since according to FATF, the objective is to report suspicious transactions, not identify the outcome of SARs. Instead, the issue seems to be a question of specifically describing the intent and purpose of a model (Canhoto, 2021; Han et al., 2020). Developing ML to improve the TM process creates a tradeoff between risk and cost (Han et al., 2020). Automating for improved efficiency in TM will result in fewer human resources, lower costs, and faster processes. On the other hand, the risk of missing SARs is a concern, as humans will no longer oversee every step of the process (Han et al., 2020). Improving effectiveness creates a tradeoff of time where the process will be lengthier and more costly but with a human-in-the-loop to interpret added information from ML methods such as graph analytics and link analysis (Han et al., 2020).

## 2.3. Explanations in high-stakes decisions

Prior research has approached the concept of a "good explanation" as a human-agent interaction problem (Miller, 2019), wherein context is essential to the person interacting with the explanation. Miller explains that we are only interested in a subset of relevant information depending on our context (Miller, 2019). Moreover, there is a growing need to understand different stakeholder requirements from ML, where xAI is evolving in response to this need (Arrieta et al., 2019). Other researchers have addressed the origins of what stakeholders want from an explanation (Brennen, 2020; P. Lipton, 2001). Originating from DARPA, the purpose of xAI was to enable human users to understand, appropriately trust, and effectively manage AI (Gunning, 2017). To address this ambitious quest, research into stakeholders' needs regarding xAI, is developing a common understanding that no one size fits all. Though, there is still a dominance of xAI built by and for data science elicit groups. However, HCI (Human-Computer-Interface) and socio-technical angles are emerging to evaluate the potential of xAI in application (Gerlings et al., 2021; Miller, 2019). Common struggles within explainability are seen as dilemmas of completeness vs. interpretability (Alvarez-Melis & Jaakkola, 2018; Gilpin et al., 2018; Z. C. Lipton, 2018) wherein generating or extracting information from models might be useful to stakeholders despite not generating a complete picture of the model. Trust (Adadi & Berrada, 2018; Shin, 2021) is another highly debated issue, with xAI aiming to enhance trust in AI on the part of end-users. Moreover, xAI is a response to goal alignment from problem formulation to model output and applicability creating biases in decision-making (if information is not used as intended) (Páez, 2019). Global model explanations, general rule extraction, additive models, and interpretability designs have all been discussed to concur implementations that deviate from the intended use (concept drift), which can cause great chaos for the affected individuals (Adadi & Berrada, 2018; Doshi-Velez & Kim, 2017). Additionally, xAI might even add to the overall model complexity, demonstrating that xAI is not always the right solution (Rudin, 2019). Choosing an xAI solution is, among other things, dependent on the people's illiteracy in the fields of both ML and the application field where the model is implemented (Doshi-Velez & Kim, 2017; Gerlings et al., 2021; Gilpin et al., 2018). Recently, the demand for explainability has increased, following the use of ML in sensitive high-risk areas such as healthcare and finance. Furthermore, the European Union has proposed regulation of AI which also puts heavy weight on explainability (De-Arteaga et al., 2019; Doshi-Velez & Kim, 2017; Zytek et al., 2021). More pragmatic approaches have surfaced to address the 'best' fit of model explanation and understandability (Páez, 2019), few of which have addressed contextual factors such as how much a ML model is automating. Zytek et al. (2021) address the influence of the degree of automation a model has on the existing workflow to understand the context of user needs and address the challenges of trust and confusion with SHAP (SHapley Additive exPlanations) (Lundberg & Lee, 2017). However, they also recognize the pitfall of displaying too much information and using terminology from data science that the intended user of the explanation may not be familiar with (Zytek et al., 2021). Specifically for high-stakes decisions, Rudin (2019) argues that explainable models do not make sense as they leave out information that the decision-maker normally incorporate when making a decision (Rudin, 2019).

## 3. Research Methods

This empirical case study is part of a larger action research study still ongoing, in collaboration with a European bank. Interviews we conducted in 2020 and 2021 along with observations of the development and testing of TMM. The aim of the study is to investigate the influence ML has on high-stakes decision-making by key stakeholders. This study forms its qualitative rigor through inductive research for the purpose of generating new theory and discovering new concepts, where the interpretivist stance supports the idea of gaining a deep understanding of the social actors involved in the phenomenon (Gioia et al., 2013).

### 3.1. Data collection

Data has been collected by the first author through semi-structured interviews, meetings, observations, daily stand-ups, e-mails, and chat messages, all of which laid the foundation for the understanding of workflows, the inner workings of TMM, and the daily operations of key stakeholders. In total, 23 interviews were conducted in the period from August 2020 to August 2021 with key stakeholders in the development department, model validation, transaction monitoring with investigation, risk assessment and management.

### 3.2. Data Analysis

The interviews were analyzed and coded through open coding and axial coding to identify the 1st order information centric dimensions derived directly from interview quotes, 2nd order labelling themes derived from the analysis, and lastly, the 3rd aggregated dimensions were based on grouped themes (Corbin & Strauss, 2012; Gioia et al., 2013). In an iterative manner, we utilized the interpretive 'insider-outsider' method (Gioia et al., 2013) by establishing an insider perspective from the first authors' 'insider view' of how automation and augmentation changes the need for explanations amongst stakeholders. The second author approached the data after collection with an 'outsider' perspective on both the phenomenon itself as well as the case setting, allowing for new ways of theorizing and identifying alternative patterns in the data. Thereafter, we initiated a more abstract 'outsider' level of analysis in order to generate a link between the 1st order codes from the interviewees and 2nd order theoretical concepts. Concepts were finally distilled into the 3rd order aggregated dimensions presenting the discovered grounds for new theory development (Benbasat et al., 1987; Gioia et al., 2013). The final data structure entails the emergence of three 3rd-order themes when implementing ML in high-risk decision-making processes: *Efficient optimization* (the underlying elements of understanding TMM (A) in general, the validation and reliability of its performance, and the debate of providing further explanations for TMM), *Diverse explanations* for taking into account the effect more or no information has on key stakeholders' work, and *Effective optimization* as the biases in augmented decision-making, revealing a need for more information on the TMM (B) output.

## 4. Case

The financial institution in scope follows the industry standards presented in Jullum et al., Han et al. and Bertrand et al. (Bertrand et al., 2021; Han et al., 2020; Jullum et al., 2020) for the AML practices of manually handling scenario-based alerts. Currently, a widely regarded problem across the sector is the fact that over 90% of all alerts generated are false positives (Han et al., 2020), meaning investigators spend their time going through an exceedingly high number of false-positive (FP) alerts. In 2018, The bank initiated a ML project to address the challenges of managing the increasingly large number of FP alerts, with a focus on critical alerts and handling cases in a timely manner. The project has been tested in a monitored environment over the past few years as the TMM has undergone iterations of improvements and feedback from key stakeholders. In 2020, the bank initiated a research initiative to identify the information needed from key stakeholders of TMM and to begin a transition toward AI-based AML approaches. The aim of TMM is to optimize manual investigation time for human investigators and minimize the time spent investigating FP alerts generated by scenarios. The time spent on investigating alerts and eventually reporting a SAR is referred to as *time-to-SAR*, which is essential for banks to remain compliant with regulations. Currently the bank is testing a set-up wherein the transaction monitoring process starts with the incoming transactions being sent through 70-100 rule-based scenarios. Only transactions that fall within the scope of the scenarios generate alerts. Alerts are then saved to a database for investigation. Here, the random forest-based TMM analyzes alerts from specifically selected scenarios with less complexity than others. These alerts are provided with a risk score (unusualness) of 1-100, which is generated by TMM. Next, the bank has tested two approaches:

**A)** Lowest-scoring alerts (FP), within the bank's risk appetite are auto-closed. As a result, investigators do not encounter these alerts, as they are to them closed. However, a dedicated team evaluates the alerts closed by TMM.

**B)** The risk score is used by capacity planners to prioritize alerts deemed high-risk. Score is visible in the interface for capacity planners and investigators.

### 4.1. The scope and value of Transaction Monitoring Model

Many different factors come into play when determining the risk tolerance in TM. In this case, the bank has decided to maintain the essential scenarios in the process around the ML model. This decision was made to accommodate the risk appetite of the bank and ensure minimal obstruction of the alert investigation. Internal experiments have determined a positive impact on the alert handling, but some challenges were also observed during testing. As one investigator put it: "*I think that the main risk is that we have to reinvestigate [alerts]in case investigations that have been closed by TMM...On the other hand, TMM basically saves us a lot of resources when it comes to alert investigation in the first part of the process and helps us to mitigate the risk of us having too many people [...] during the first 10-12 days and idle for the rest of the month, because there will be no work for them. So, TMM takes care of this 'over capacity' issue that we might have.* (Investigator). Based on the results of internal experiments, the bank wanted to further explore the effect of the ML model in the transaction monitoring framework in order to better understand complications and how to improve the unified performance of human and machine.

### 4.2. The role of TMM key stakeholders

The key stakeholder groups in this research are defined by working in the near environment of TMM and being either indirectly or directly affected by TMM in the decision making and work around TM.

**Capacity planners,** also referred to as **Coordinators**, plan the production of incoming alerts with the optimal number of employees. Alerts are produced in batches. For each batch, they estimate the number of alerts across scenarios and the necessary resources. Coordinators often have a background as investigators themselves and are now senior specialists with knowledge in specific fields, acting as a point of contact for investigators if they are doubting a decision. Furthermore, their role is to keep track of production (how many alerts are escalated/closed per investigator and per scenario). Coordinators prioritize and distribute alerts based on scenario types and time (urgency). TMM is set up to run through the batches of alerts, generate the risk scores, and/or auto-close the lowest-scoring alerts. When closing alerts (A), the model reduces the workload of investigators by ~20%. Coordinators are affected by the model's closing performance in their planning of production as the model alleviates the workload on more simple cases and thereby ensure more high-risk alerts get addressed within the required timeframe. Moreover, TMM (A) reduces the resources needed for alert handling. When augmenting (B), capacity planners utilize the risk scores to prioritize and distribute more suspicious alerts first to minimize the time-to-SAR ratio.

**Investigators** work in teams, oftentimes centered around specific types of scenarios or local regulatory knowledge. Their role as investigators is sometimes split, so they have other responsibilities as well, such as specialist roles, scenario development or risk evaluations, depending on their seniority and skillset. Therefore, capacity planners need to know which investigators they need for an upcoming batch of alerts and when they need them. Investigators each get a list of alerts from their team coordinator, which they are expected to handle within a given timeframe to ensure compliance and time for extended alert investigation for alerts sent further on to case investigation. Case investigation takes place in more advanced teams, where multiple alerts can be compiled into one case investigation. If cases are not closed, they are eventually reported as SARs to authorities. The timeframe investigators operate within is very tight and can have tremendous consequences for the bank if they fail to comply, so TMM's reduction of FP alerts was welcomed. When TMM (A) closes alerts, it does not directly impact the workflow for the investigators, as they never see the closed alerts, but only sense the reduction in their workload and the increased complexity of the remaining alerts they do handle.

**Model Validation Unit.** An internal model validation team is actively working with the development team in the development and testing of approaches for TMM. The purpose of the model validation team is to validate all risk-related models in the bank. Together with the model owner at the management level and a risk committee, the departments establish the desired risk appetite for the TMM. Thereafter, it is the model validation unit's task to validate the model based on the logic behind the problem formulation to the model selection, set-up, feature engineering, implementation, and monitoring.

**Development Team.** The development team includes specialists from the industry who understand the TM process, regulations and risk assessments needed, in addition to data scientists, full-stack developers, architects and physicists who are working on developing and improving TMM. During the development stages, they have been testing different approaches, models and set-ups to ensure the best reliability and working performance under the given circumstances. Moreover, they have developed a monitoring model with extended information on model performance and data consistency. This allows the team to identify any inconsistency in data flow or outliers in predictions. The last key stakeholder group is a combination of **internal auditors and external regulators,** such as the FSA (financial supervisory authority). These stakeholders take client samples and inquire about anything from how the AML process is working to the explanation behind a single alert outcome. Their questions must be answered to stay compliant.

## 5. Findings

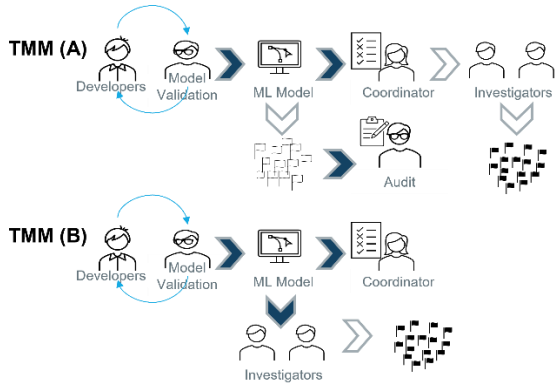

Figure 1: TMM (A) and TMM (B)

The analysis of the interviews showed a divergence in explainability needs stemming from the underlying shift between the objectives of automation: TMM (A) – reducing alert volume and augmentation: TMM (B) – reducing time-to-SAR while preserving the end-to-end process goal of detecting financial crime. The following section will present in further detail the essential differences in terms of information need between the two approaches, starting with automation of closing low-risk alerts.

### 5.1. Automation of low-scoring alerts

The purpose of the first model TMM (A), is to remove unproductive work for the investigators and make them spend less time on FP alerts, illustrated in the Figure 1: TMM (A) TMM(A). The key stakeholders interacting with the model, from left to right, illustrate how model validation and developers work together to develop a sound model. The darkened arrows emphasize increased interest from key stakeholders in knowing details about how the model operates and decides whether to close/not close an alert. To ensure model performance, the development team has developed a monitoring framework displaying various performance metrics and data statistics for their own use. This displays data for test comparison, data drift and statistics from alerts closed by TMM. The monitoring framework is mainly intended to generate information for the developers to rely on and validate the scope of closed alerts. However, for now, auto-closed alerts are checked by dedicated investigators and risk owners to ensure the model is operating according to the agreed risk appetite. *"During the period of testing, there is a test report that tells us if it is actually performing as expected or not."* (Developer). On the other side of the model in the figure, the coordinator receives statistics on closed alerts per batch, which they use to predict their upcoming planning and production of batches. However, they are not provided with extended information on the model's performance, accuracy, or ratios to give them some context on the model's performance. At the same time, investigators are taken out of the information loop, as they are not interacting directly with the automation model and should be able to form their own decisions on the prioritized human alert investigation. *"...the current setup is meant for the model to be completely invisible to investigators. This is to reduce the volume. It doesn't really matter whether investigators understand it or not, despite us having many talks with them saying that they would like to understand it more. What we focus on now is decreased volume."* (Data Scientist). Moreover, investigators are not accountable for the auto-closed alerts made by TMM. *"The biggest problem for [Investigators] is not correct decisions made by software. It's only correct decisions made by humans, right? So, they have their own policies and training. We have extremely good capabilities for this, so the main concern of their department is FTEs"* (Data Scientist). This highlights the effect (or the lack thereof) it has on investigators, leaving them to focus on more complex alerts that have not been closed by TMM. However, it also leaves the reliability of auto-closed alerts to the development team for the closing of alerts. Alerts closed by investigators have a written closing comment from the investigator describing the outcome of the individual alert. In contrast, alerts closed by TMM are accompanied by the model risk score. The explanation for closing an alert is written in natural language, whereas the score generated by TMM is open to interpretation and does not provide a sufficient explanation for auditors or case investigators who might have to include auto-closed alerts later in the investigation phases. This can eventually become an issue or a source of additional work if an alert becomes a part of an internal audit or the FSA requires information on a case where alerts have been auto-closed by TMM. When asked about what information is important during audits and alert investigation, a capacity planner answered: *"All the reasoning behind actions taken with alerts is important. We might as well be asked for data analysis and then, for example, look into statistics of how many others get generated... For example, the TMM-based closure ratios based on scenarios"* (Capacity Planner). Requesting the reasoning behind the closure of alerts is understandable as the current score does not provide any relatable information on the close, and capacity planners become responsible for the entire production of alert investigation for a specific country. The risk score from TMM (B) and metrics from the monitoring will not be able to satisfy the auditors' or regulators' requirements in these cases. However, a full explanation might not be possible due to the complex AML environment the model operates within. *"It's also a risk of explainability. It's that the environment we have this model in is quite complex. I think that is a risk, that we might not be able to fully explain why the model auto-closed some alerts and didn't close other alerts. I don't think anyone can really say OK it closed because of this and that. That's a risk."* (Data Scientist). In the case of complexity in ML models, it is in the eyes of the beholder, where different stakeholders having their own prerequisites for understanding and interpreting the model results. The information required for an explanation designed to the developers lies in the monitoring framework, with metrics ranging from statistical information on data and eventual imbalances on features to ROC curves and confusion matrices. The information requested by the coordinators and investigators is centered around business rules and their everyday context, which relates to their specialty. This becomes clear when discussing the pros and cons of manual and automated alert handling. *"…manual investigation makes sense as it will give a broader picture of the customers and not just the transactions that TMM is probably looking into. So where is the customer living? What is the pattern of the customers, which stores do the customers go into…"* (Investigator). Though unsure of what data TMM is operating with, this underlines the information the investigator uses to identify a suspicious transaction, whereof not all data is available for TMM. From a data scientist perspective, an explanation for why an alert was closed might not be in line with what investigators are used to seeing. *"It's an alert closed. We have an explanation that would be good for auditing purposes in case someone asks us: why did the model close this? But it wouldn't be useful for investigators."* (Data Scientist). Acknowledging the differences between the fields of AML and ML, this quote from a data scientist is consistent with the following quote from a developer asked if investigators would benefit from knowing more about TMM. *"So, should the investigators know more about it? I don't know. Precision, accuracy scores, ROC curves, or instance-based explanations? …I don't think that will be very informative. So, there is definitely the chance to bring something understandable."* (Developer). Acknowledging that the usual performance metrics used in ML will be of no use to the investigators, the developer also sees the possibility of creating something that would be understandable from investigators point of view.

## 5.2. Augmentation of alert prioritization

To accommodate challenges around information need from the auto-closing model, an augmentation approach was tested. TMM (B) made the risk score available to capacity planners and investigators and shifted the responsible party of the final decision to close or escalate an alert. Now, alerts are not closed but given a score of high, medium, or low risk, leaving investigators to make the final decision. The change in objective leaves a risk of biased decisions emerging and a remaining need for further information on model performance persists. As shown in Figure 1, investigators receive information directly from the model, which they convey in their alert investigation. *"The thing is…I never actually looked specifically at these numbers, but I know that alerts have this low, medium, high priority. This influences me quite a bit… if I see a high priority, then I think this is something that I should take into consideration. So, yes, knowing about this scale absolutely influences the way I would treat an alert …"* (Investigator). Another investigator and risk specialist share this concern but also point out the duality of the situation: *"This might give some indication to the investigator that there's a high chance that this alert will be closed. I'm just wondering if it might be good. It might help some people with the decision. On the other hand, it might create a mental bias toward the decision generated by the investigator themselves. So, like, every stick has two ends, right?"* (Investigator). To address this bias, capacity planners pitch in and argue for more information about why the TMM scores are the way they are, both for the sake of the investigators and the capacity planners themselves. *"I don't know the meaning of the scores, so it was maybe mentioned somewhere, but I don't recall such documentation. So, the low score. What does it mean?…[]…with capacity planners, I think it will be good to know [more]. They're actually always asking about the numbers, why they are so low, for example, and they will also understand the investigation point of view"* (Capacity Planner). Furthermore, a concern over the time spent investigating the alert score emerges, as well as a desire to understand why TMM has produced a high or low score: *"…but when it comes to the ones that TMM is saying are really high-risk… further information can direct the investigator into why TMM has given it a high score. I think that would be good, because otherwise there is a risk that the [investigators] spend time scratching their heads over why TMM gave a high score"* (Capacity Planner). On the other hand, capacity planners use the risk scores from TMM to prioritize incoming alerts, so the ones with the highest scores are handled first in an effort to decrease time-to-SAR. When informed about an upcoming update to the interface that might exclude the TMM (B) risk score, a capacity planner expressed its importance for their work. *"No, it is important! Don't take it away. I don't have the power to decide anything ...but if we see that some scenario is producing more [potential] SARs, then we need to take them first instead of those that are creating zero SARs. It is beneficial to know that."* (Capacity Planner). In conclusion, investigators' access to the current information generated by TMM (B) highlights a potential risk of the investigators being negatively affected and biased by the alerts score due to investigation bias or unproductive time spent trying to understand why a certain score was given. This suggests that hiding the score or removing the model scoring in favor of a more informative explanation for investigators, such as LIME or general feature importance methods, should be evaluated further. As the risk score is currently disclosed to both capacity planners and investigators, a strong tendency towards requiring more information was apparent in both stakeholder groups. Moreover, the need for showing the model reasoning to audit remained in moth cases.

## 6. Discussion

The divergence between the model objective of automation and augmentation approach to optimizing the TM process has shown to cause various concerns regarding key stakeholders' interaction with TMM. From the analysis, a few specific areas stand out when differentiating between automation and augmentation. First, contrary to why many studies argue for xAI, no lack of trust (Ribeiro et al., 2016; Shin, 2021) or algorithm aversion (Burton et al., 2019; Lee, 2018) was discovered for investigators or capacity planners when analyzing both versions of TMM. This may be due to the reduction of FP without their intervention and decreasing the time-to-SAR by prioritization of alerts with no influence on their work. Moreover, it is only when audits occur or when case investigators need to re-open alerts during case investigation that more information from TMM (A) is required. Here, the information goal is to justify or verify the decision for auditors (Arrieta et al., 2019; Samek et al., 2017) and explain the "why" to case investigators, which is seen as more challenging as they require more contextual information in language they can relate to from their work (Arrieta et al., 2019; Doshi-Velez & Kim, 2017; Miller, 2019; Zytek et al., 2021). The evaluation metrics capacity planners and investigators ask for weigh more heavily on the interpretability end of the completeness-interpretability scale (Gilpin et al., 2018). Hence, using the monitoring framework to deliver information and explanations would potentially impose a risk of 'inmates running the asylum' (Miller et al., 2017) as the context of the information is far from what they are familiar with and would potentially increase time spent figuring out what the information and numbers represent. Hence, TMM (A) causes the investigators and, to some degree, capacity planners to perform their work without much concern for being liable for the model's performance; the liability here lies in the hands of the developers and data scientists. Holding the data scientists accountable when automating the auto-closing of alerts enables them to utilize more advanced and closer-to-completeness tools for explainability together with the monitoring framework when ensuring a properly performing model. Here, explainability methods such as PDP (Partial Dependency Plots) (Goldstein et al., 2015) and SHAP (Lundberg & Lee, 2017) can be utilized for exploration and improvement of the model (Samek et al., 2017). Changing the scope from automation to augmentation of TMM may not seem that significant, but it shifts the liability of closing alerts back to the investigators, who now face the challenge of interpreting the model output, as it is visible in their interface. A crucial finding here is the risk of biased investigation based on the risk score produced by TMM (B) and the risk of unproductive time spent on understanding the score in its current form. Therefore, the information requirements and needs that were previously reserved for auditors and case investigators are now required for investigators as well. Otherwise, the score could be removed altogether from the investigators' interface, to avoid model biases. However, both capacity planners and investigators articulate the benefits of knowing more about the score, as it can be used to point the investigator in the right direction during an investigation and reduce time-to-SAR.

## 7. Conclusion

The financial sector is under significant pressure to optimize their AML practices, specifically TM. Due to the outdated, labor-intensive process of reviewing alerts, a European bank has initiated a project for optimizing the process with machine learning. Two designs were built based on a random forest model and utilized for automation (A) and augmentation (B). As part of the project, this action research was initiated to explore which information was required by different stakeholders to ease the shift towards machine learning-based TM. It was found that depending on the design and objective of the models tested, explanation and information needs varied among key stakeholders, which argues for building stakeholder-specific explanations not only focusing on the end-user. In TMM (A) the objective of which was to auto-close low-scoring alerts, investigators were not affected by the model and had a limited need for information. Moreover, they did not display any trust-related concerns regarding the model (A), which could be because of their limited involvement and the positive effect it has on their work. Capacity planners were able to reduce the resources needed for handling incoming alert batches and reduce unproductive work. However, when audits occurred or case investigations were extended, these capacity planners required reasoning and validation for why alerts had been auto-closed. This type of explanation was articulated as a need for context-relatable explanations that are different from model performance metrics provided in the monitoring framework. Therefore, explanations for capacity planners should focus on interpretability rather than completeness, to generate understanding and trust. The objective of TMM (B) is to score incoming alerts for prioritization purposes and thereby minimize time-to-SAR but risk increasing it without contextual explanations. While developers were once reliable for the auto-closed alerts, while capable of interpreting the monitoring framework, they now provide a score to capacity planners for prioritizing production. Since the scoring is also visible to the investigators, it creates a risk of biased investigation. If the score is low, effort is low and vice versa. Additionally, the risk of investigators spending time on interpreting the score and figuring out why it is the way it is adds to unproductive work and increase time-to-SAR contrary to the bank's objective. To solve these issues, the score could either be removed from the investigator's view, or more contextual explanations could be added to point the investigator in the right direction of the issue with an alert. Moreover, auditors may still require information on the model logic and its operation for compliance, which argues for exploring global explanations. Contributions from this paper add to the sparse literature on machine learning in AML, especially in TM (Canhoto, 2021), adding a socio-technical perspective on an empirical case. Moreover, the paper sheds lights on different stakeholder needs as the objectives of machine learning models changes. It also proposes the need to address requirements for xAI in automation and augmentation processes differently as the liable party of the outcome shifts and is not always the end user. The paper responds to the call for more interdisciplinary work in explainable AI and identifies the stakeholder requirements for building xAI. For future work, there is a demonstrated need for further empirical research on the shift in stakeholder information and explanation needs, depending on automation or augmentation. Moreover, empirical research testing xAI frameworks to satisfy stakeholders needs and concerns are called for.